\documentclass[sigconf, screen]{acmart}

\AtBeginDocument{%
  \providecommand\BibTeX{{%
    \normalfont B\kern-0.5em{\scshape i\kern-0.25em b}\kern-0.8em\TeX}}}

\copyrightyear{2024}
\acmYear{2024}
\setcopyright{acmlicensed}\acmConference[SIGIR '24]{Proceedings of the 47th International ACM SIGIR Conference on Research and Development in Information Retrieval}{July 14--18, 2024}{Washington, DC, USA}
\acmBooktitle{Proceedings of the 47th International ACM SIGIR Conference on Research and Development in Information Retrieval (SIGIR '24), July 14--18, 2024, Washington, DC, USA}
\acmDOI{10.1145/3626772.3661368}
\acmISBN{979-8-4007-0431-4/24/07}

\usepackage{balance}
\usepackage{caption}
\usepackage{subcaption}
\usepackage{soul}
\usepackage[inline]{enumitem}
\usepackage[normalem]{ulem}
\usepackage{multirow}

\begin{document}

\newcommand{\srijan}[1]{\textcolor{blue}{[srijan: #1]}}
\newcommand{\oj}[1]{\textcolor{red}{[OJ: #1]}}

\newcommand{\E}{\mathbf{E}}
\newcommand{\x}{\mathbf{x}}
\newcommand{\y}{\mathbf{y}}

\title{Monitoring the Evolution of Behavioural Embeddings in Social Media Recommendation}

\author{Srijan Saket}
\email{srijanskt@gmail.com}
\affiliation{
    \institution{ShareChat}
    \city{Seattle}
    \country{US}
}
\author{Olivier Jeunen}
\email{jeunen@sharechat.co}
\affiliation{
    \institution{ShareChat}
    \city{Edinburgh}
    \country{UK}
}

\author{Md. Danish Kalim}
\email{danish@sharechat.co}
\affiliation{
    \institution{ShareChat}
    \city{Bangalore}
    \country{India}
}

\makeatletter
\let\@authorsaddresses\@empty
\makeatother

\begin{abstract}

Emerging short-video platforms like TikTok, Instagram Reels, and ShareChat present unique challenges for recommender systems, primarily originating from a continuous stream of new content. ShareChat alone receives approximately 2 million pieces of fresh content daily, complicating efforts to assess quality, learn effective latent representations, and accurately match content with the appropriate user base, especially given limited user feedback.
Embedding-based approaches are a popular choice for industrial recommender systems because they can learn low-dimensional representations of items, leading to effective recommendation that can easily scale to millions of items and users.

Our work characterizes the evolution of such embeddings in short-video recommendation systems, comparing the effect of batch and real-time updates to content embeddings.
We investigate \emph{how} embeddings change with subsequent updates, explore the relationship between embeddings and popularity bias, and highlight their impact on user engagement metrics. Our study unveils the contrast in the number of interactions needed to achieve mature embeddings in a batch learning setup versus a real-time one, identifies the point of highest information updates, and explores the distribution of $\ell_2$-norms across the two competing learning modes. Utilizing a production system deployed on a large-scale short-video app with over 180 million users, our findings offer insights into designing effective recommendation systems and enhancing user satisfaction and engagement in short-video applications.

\end{abstract}



\maketitle

\section{Introduction}
Short-video applications have experienced significant growth in recent years, with users seeking entertainment in short-form engaging content.
With the advent of these applications has come a surge in specialised research literature aimed at adapting and improving recommender systems specifically for short-video recommendation~\cite{Liu2019,Pan2023,Gong2022,Liu2023,Jeunen2023_RecSys,Jeunen2024_FIRE}.
Embedding-based approaches have long been a staple in industrial recommender systems, and this is no different in short-video applications.
Such systems leverage models like Field-aware Factorization Machines (FFMs), wide-and-deep neural networks, and two-tower models to learn vector representations for users and content, often referred to as \emph{embeddings}~\cite{cheng2016wide, zhang2019deep, guo2017deepfm}.
The unique nature of short-video content presents some additional challenges.
With constant new content uploads and a vast content space, recommending relevant videos becomes complex, especially for new uploads for which we cannot rely on historical user interaction data.
Over time, the system can usually learn behavioral embeddings to capture user preferences.
Nevertheless, this assumes that all content receives a reasonable amount of impressions, which might not always be the case.
As such, existing challenges in the field of recommendation are exacerbated with short-video content, such as the requirement for diversity, varying shelf-life of items, and general cold start problems, as mentioned in~\cite{memer, hara2018can, bojanowski2017enriching, hershey2017cnn}.

To address these issues, it becomes essential to gain a comprehensive understanding of how embeddings \emph{evolve} as they are updated with new user interactions over time.
By studying the evolution of embeddings, we can uncover valuable insights into content dynamics and improve the evaluation and effectiveness of recommendation algorithms.
In this work, we focus on characterizing the evolution of learned embeddings of short-video content and propose metrics that enable us to gauge both the quality of the item embeddings themselves, and the recommendations they generate.
To understand how item embeddings evolve, we consider two common modes of learning used to update embeddings:
\begin{enumerate*}[label=(\roman*)]
    \item \emph{batch} updates, where user interactions are accumulated and embeddings are updated in batches every $X$ hours, and
    \item \emph{real-time} updates, where the embedding for every item is updated in real time when new user interactions occur.
\end{enumerate*}

\begin{figure}[t]
  \centering
  \includegraphics[width=8.5cm, height=2.5cm] {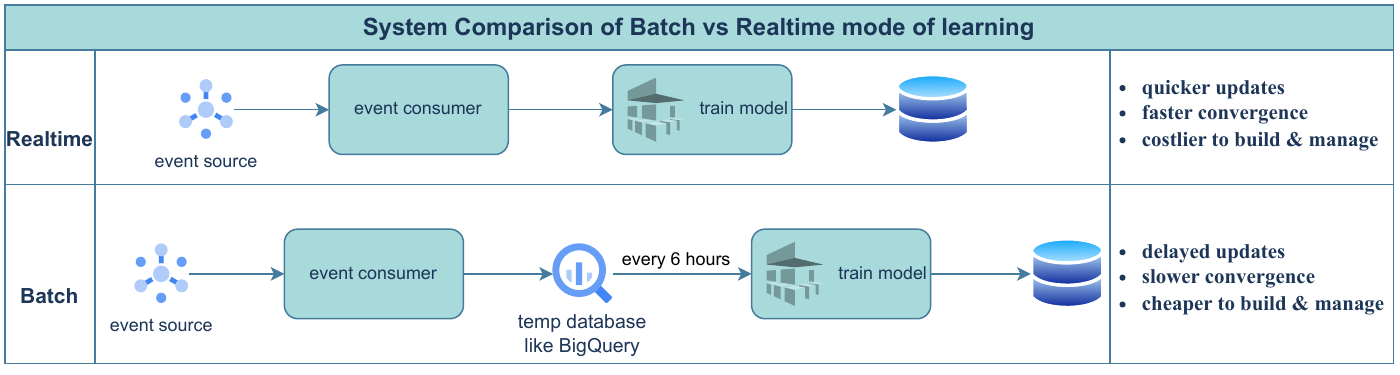}
  \caption{System comparison of real-time vs batch learning for content embeddings.}
  \label{fig:system_comparison}
  \vspace{-2pt}
\end{figure}

Figure~\ref{fig:system_comparison} highlights the key differences of this engineering choice, that can have a significant impact on the quality of the recommender system overall.
For both learning modes, we characterize the evolution of embeddings along a number of aspects.
First, we focus on embedding convergence.
When the embedding for an item has converged, we say it has ``\emph{matured}''. We define embedding \emph{maturity} as the stage where incremental updates yield minimal changes in the embeddings. This is characterized by $1-\cos({\rm emb}_t, {\rm emb}_{\rm converged})< \alpha$, where ${\rm emb}_{\rm converged}$ is the embedding value after convergence, ${\rm emb}_t$ is the value at time $t$ (or after $x$ views) and $\alpha$ is determined based on the specific business use case, such as item recommendation performance.
We then consider the question: ``how many user interactions does an item require to obtain a \emph{mature} embedding?''
We find that real-time embeddings mature significantly faster, requiring only 20\% of the number of user interactions that are required by batch learning modes.

Second, we study the trajectory of embeddings over time by comparing the cosine similarity of an embedding at time $t$ and the subsequent timestep $t+1$.
We use this estimate to quantify the \emph{information} of embedding updates that occur with each new user interaction.
We note that the highest information updates occur notably earlier in real-time learning compared to batch learning.
Consequently, real-time training leads to a quicker convergence of embeddings towards their final vectors compared to batch training. 
Furthermore, we cross-validate this insight with user engagement metrics including video click-rate and ``successful video plays'' across view buckets, which supports our empirical observations.

Third, noting that most embedding-based recommenders rely on the $\ell_2$-norm of embeddings, we investigate how the $\ell_2$-norm is distributed conditional on impression counts, for both real-time and batch learnt embeddings.
We observe that batch-trained embeddings suffer from significantly higher norms, up to a factor $\times 5$ for high-impression posts compared to low-impression posts.
This contributes significantly towards the popularity bias in content recommendations on the platform.

Our work presents a detailed analysis of two common learning modes (i.e. batch or real-time), for embedding-based recommender systems deployed at scale on a short-video app serving recommendations to over 180 million users.
It is our hope that the analysis and findings presented in this article can inform the design of content recommendation systems for other practitioners, especially at the early stages of new content being uploaded.

\begin{table}[h]
\centering
\begin{tabular}{ccrr}
    \toprule
    \textbf{Update Type} &	\textbf{Signal Type} & \textbf{\# Posts} & \textbf{\# Updates} \\
    \midrule
    \multirow{5}{*}{Batch} &	click & $\sim42$K & $\sim42$K \\
    &	share & $\sim54$K & $\sim76$K \\
    &	view & $\sim100$K & $\sim100$K \\
    &	skip & $\sim108$K & $\sim108$K \\
    &   like & $\sim120$K & $\sim177$K \\
    \hline
    \multirow{5}{*}{Realtime} & click & $\sim145$K & $\sim34$M \\
    & share & $\sim145$K & $\sim34$M \\
    & view & $\sim140$K & $\sim18$M \\
    & skip & $\sim140$K & $\sim18$M \\
    & like & $\sim145$K & $\sim34$M \\
    \bottomrule
\end{tabular}

\vspace{3mm}
\caption{Number of posts and embedding updates using the FFM model for the dataset used in this paper.}
\label{data_tab}
\end{table}

\section{Evolution of item embeddings}
In feed recommendation models, a common method to prioritize content for a user in a recommender system is by assessing the dot product between the user's embeddings and the item's embeddings \cite{shalev2014understanding, mikolov2013distributed}. Hence, the norm and the angle of embeddings in Euclidean space play a major role in deciding its ranking order in the user’s feed. We begin by describing the dataset collected and then proceed to characterize the embeddings in subsequent subsections.

\subsection{Dataset \& Embedding methods}
We gathered a dataset of video posts created in the Hindi language over a period of one week on the ShareChat application, capturing both implicit signals (such as video play, and skip) and explicit signals (including click, like, share). For each signal, we utilized Field-aware Factorization Machines (FFM) \cite{ffm_paper, juan2017field} to learn 32-dimensional embeddings from user interactions. The dataset consists of time-series data, where each post is associated with an embedding learned at different points in time. 

We focus on item embeddings for exposition and consider distance measures such as $\ell_2$-norm and cosine distance of item embeddings from different reference points (like initial value or running difference). We compute the above-mentioned metrics for two modes of learning behavioral embeddings namely, batch and real-time. The batch setup updates the embeddings at a specified frequency by collecting data over a period of \textit{six} hours whereas the real-time method dynamically updates them based on every interaction data point.

In Table \ref{data_tab}, we provide the number of posts and embedding updates associated with the dataset for each signal type. Note that the ratio of embedding updates with the number posts is much smaller for batch updates. This is because in the batch case, updates are sporadic and not done after every interaction.  

For simplicity, we focus our analysis on the ``view'' signal, although we expect similar trends for other signals since the learning method remains consistent. To ensure a sufficiently large number of interactions for capturing evolution, we retained posts with views equal to or exceeding $100\, 000$. The posts being analysed in Table \ref{data_tab} have this filter applied. Throughout this paper, the terms ``views'' and ``interactions'' are used interchangeably, indicating user engagement with the video content.

\subsection{Embedding Maturity with Views}
 
\begin{figure}[!t]
  \centering
  \includegraphics[width=8cm, height=5cm] {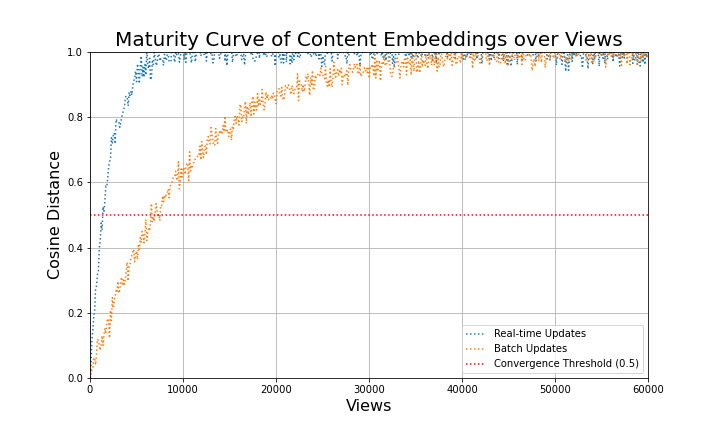}
  \caption{Comparison of item embeddings' maturity curves for batch \& realtime. Realtime achieves convergence significantly faster than batch updates do.}
  \label{fig:emb_evolution}
  \vspace{-2pt}
\end{figure}

Figiure \ref{fig:emb_evolution} illustrates the cosine distance between the content embedding at a given time $t$ and the embedding at the converged state, plotted against the number of views $x$ the content has received on the platform.

Assume we use a cosine distance threshold of $0.5$ to evaluate the maturity of the learning process for specific content. Our analysis reveals that batch embeddings surpass this threshold at approximately $10\, 000$ views, whereas realtime embeddings achieve it at around $2\, 000$ views. This suggests that utilizing realtime updates leads to considerably faster learning of embeddings compared to the batch mode.

\begin{figure}[!t]
  \centering
  \includegraphics[width=8cm, height=5cm] {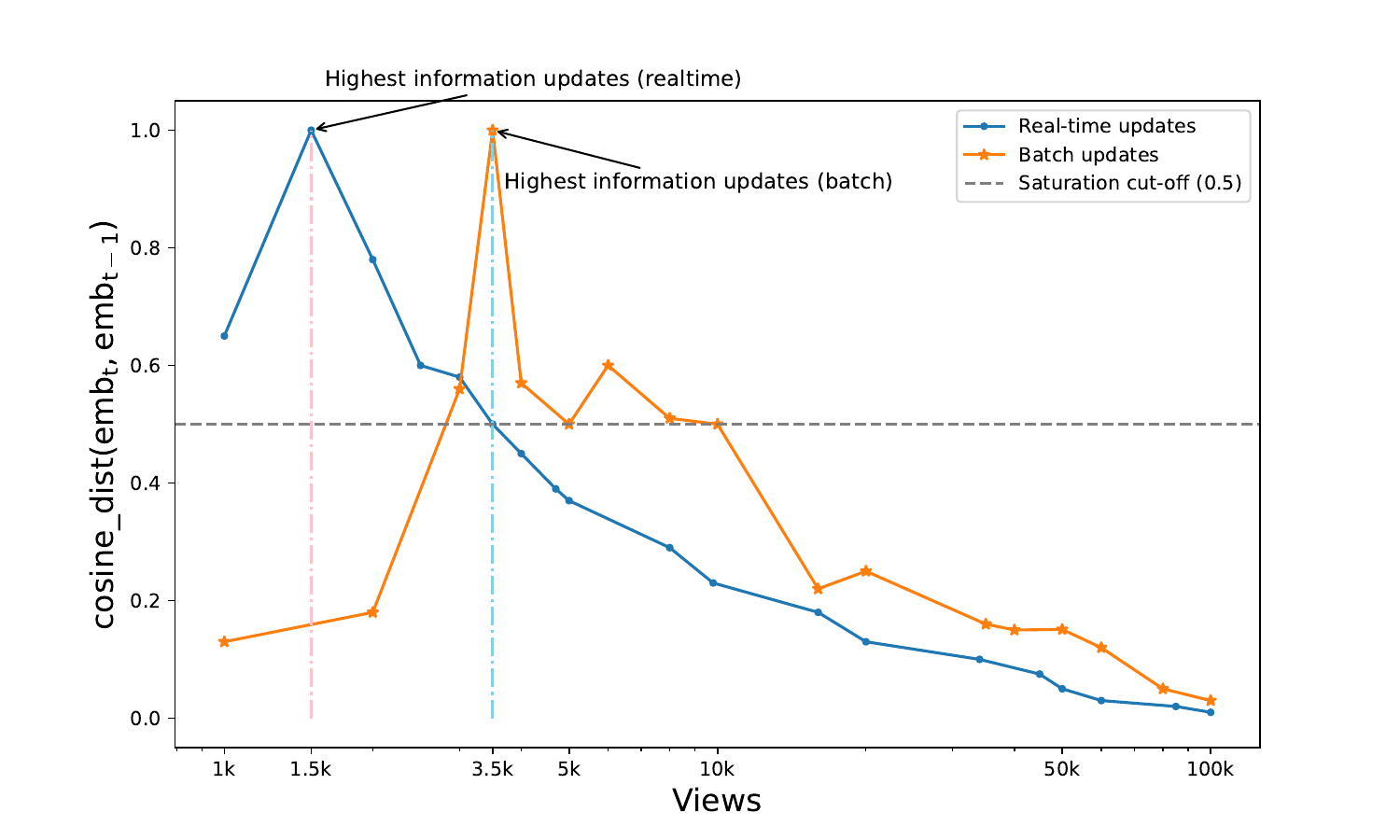}
  \caption{Comparison of item embeddings' maturity curves for batch \& realtime. Realtime achieves convergence significantly faster than batch updates do.}
  \label{fig:peak_learning_plot_realtime}
  \vspace{-2pt}
\end{figure}

\begin{figure*}[t!]
    \centering
    \begin{subfigure}[t]{0.45\textwidth}
        \centering
        \includegraphics[width=7cm, height=4.5cm]{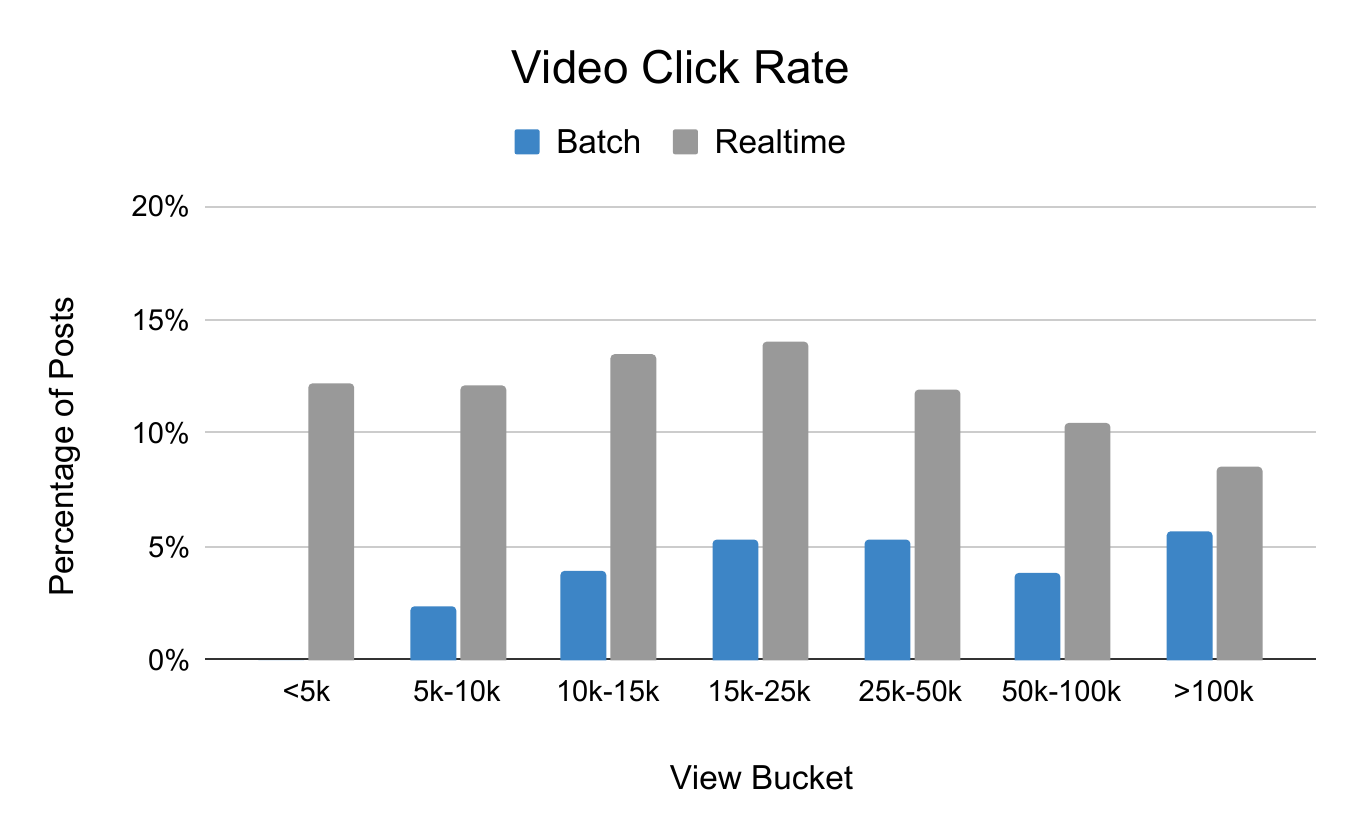}
    \end{subfigure}
    \hspace{1mm}
    \begin{subfigure}[t]{0.45\textwidth}
        \centering
        \includegraphics[width=7cm, height=4.5cm]{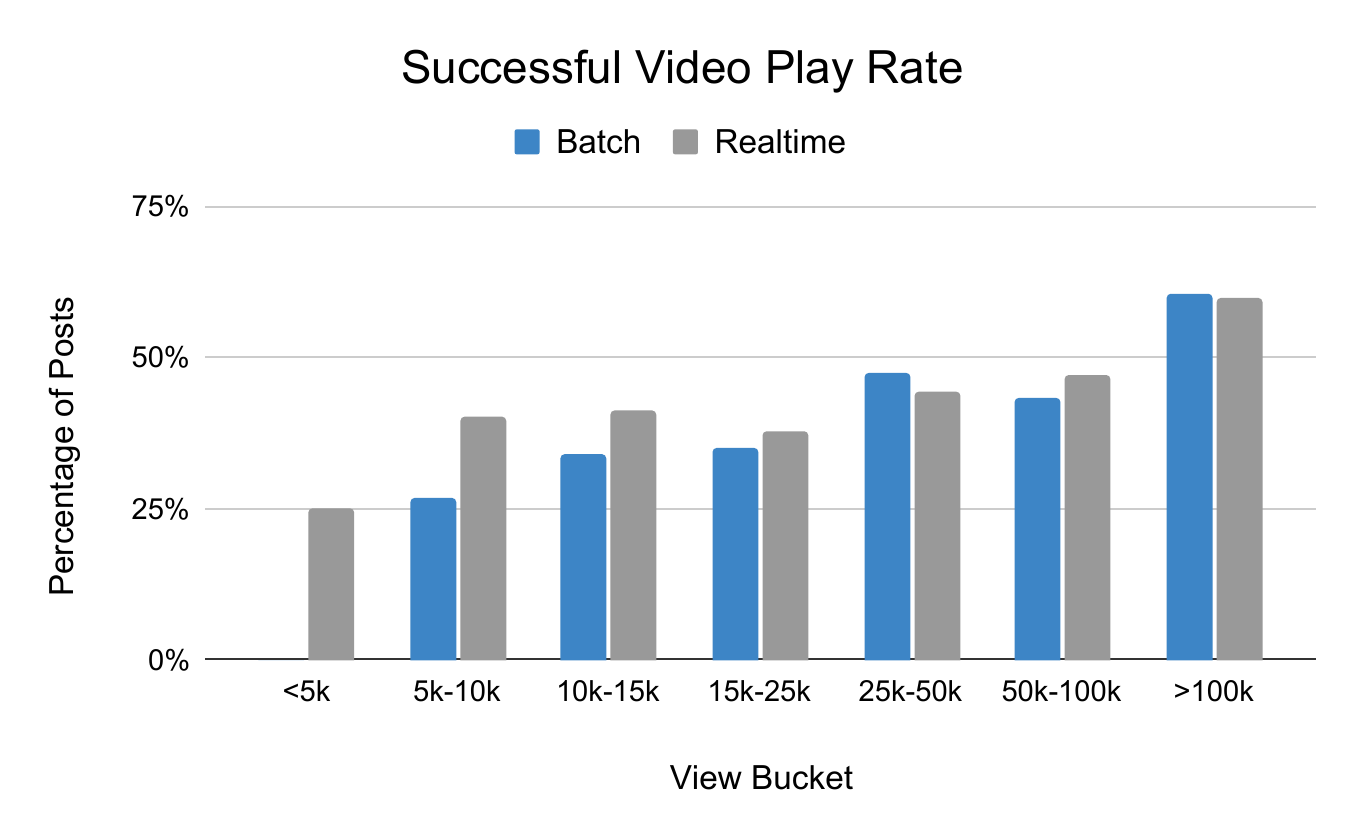}
    \end{subfigure}
    \caption{Video click rate and successful video play from experiments on the described dataset.}
    \vspace{-5mm}
\label{fig:business_metrics_ctr}
\end{figure*} 

\begin{figure*}[t!]
    \centering
    \begin{subfigure}[t]{0.45\textwidth}
        \centering
        \includegraphics[width=7cm, height=4.5cm]{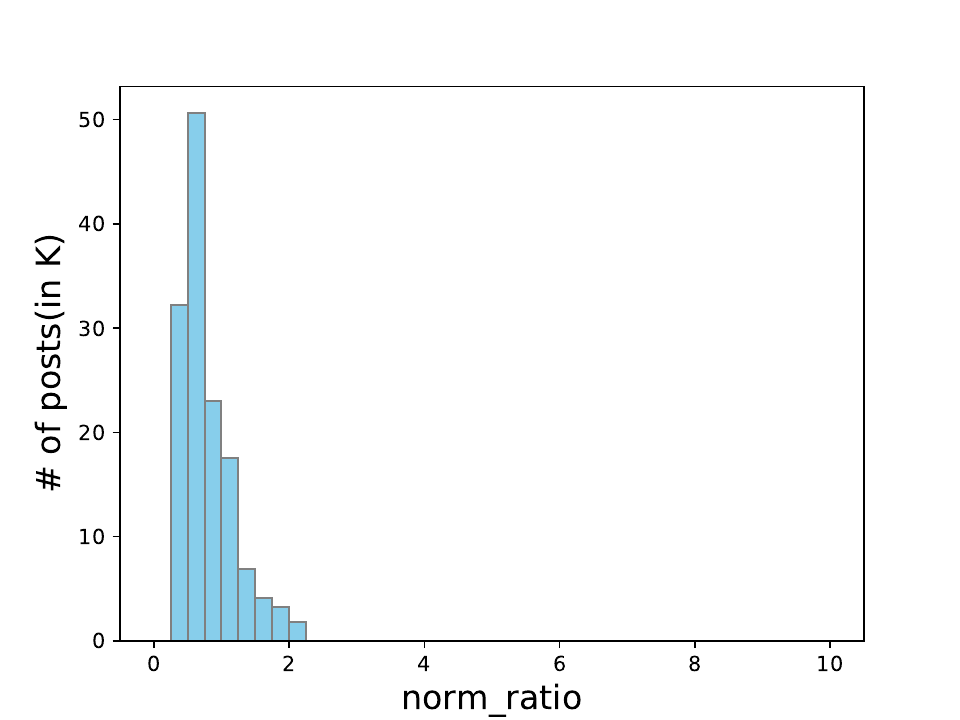}
        \caption{
        Distribution of norm ratio for realtime embedding updates
        }
        \label{fig:norm_ratio_realtime}
    \end{subfigure}
    \hspace{1mm}
    \begin{subfigure}[t]{0.45\textwidth}
        \centering
        \includegraphics[width=7cm, height=4.5cm]{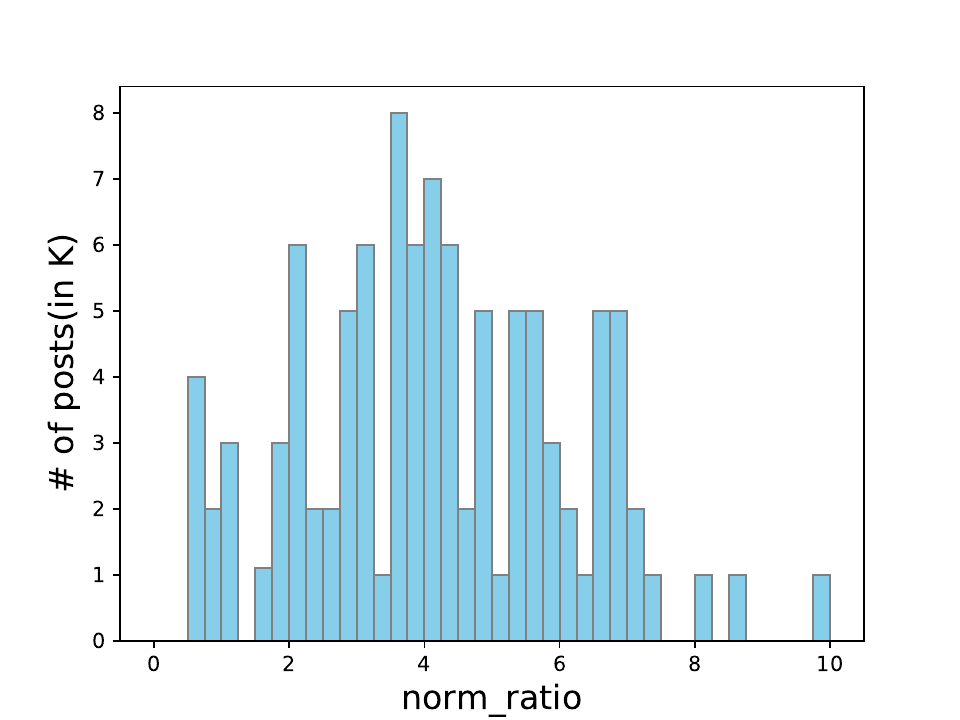}
        \caption{ 
        Distribution of norm ratio for batch embedding updates
        }
        \label{fig:norm_ratio_batch}
    \end{subfigure}
    \caption{For batched updates, the norm ratio is higher for converged embeddings w.r.t. initial embeddings.}
    \vspace{-2mm}
\label{fig:norm_ratio}
\end{figure*}

However, utilizing a fixed threshold as a means of quantifying our findings mathematically is not the most optimal approach, as this threshold is heavily influenced by various factors such as the data, model, and initialization strategy of the embeddings. Consequently, we turn our attention to examining the variation in cosine distance with each additional view the content accumulates, averaging this difference across all item embeddings. By plotting this information, we can identify the point of highest-information updates and determine the expected number of views or interactions required beyond that point to achieve convergence for both real-time and batch learning strategies. This approach allows for a more nuanced understanding of the learning process and facilitates the comparison of different learning strategies.

For this purpose, let us first define checkpoints on views as $V_1, V_2, \cdots, V_k$. Further, let us represent the embedding vectors of content at the number of views = $V_i$ as ${V_i}^1, {V_i}^2, \cdots, {V_i}^{N_i} ~ \forall ~i\in [1,k]$, where $N_i$ is the number of posts on the social media platform that reached at least $V_i$ views. Then, the amount of learning for embeddings at views $= V_i$ can be defined as

\setlength{\belowdisplayskip}{0pt} \setlength{\belowdisplayshortskip}{0pt}
\setlength{\abovedisplayskip}{0pt} \setlength{\abovedisplayshortskip}{0pt}

\begin{equation} \label{eq:1}
    L(V_i) = \frac{1}{(V_i - V_{i-1})}\sum_{j=1}^{N_i} {\rm Dist}({V_i}^j, {V_{i-1}}^j) ~~~~~~~ \forall ~ i \in [1,k],
\end{equation}
\vspace{0.5mm}

where ${\rm Dist}(x,y)$ can be any distance metric. In this paper, we use cosine distance as the metric, that is, ${\rm Dist}(x,y) = 1 - \frac{<x,y>}{\lVert x\rVert_2\lVert y\rVert_2}$ 
Since we are looking at the average across thousands of content posts, statistics like the threshold are more stable.

Figure \ref{fig:peak_learning_plot_realtime} demonstrates the model's learning progression at different checkpoints, using Eq. \ref{eq:1}. It is observed that in the real-time mode, the content embeddings reach their peak learning point twice as fast compared to the batch mode. Additionally, the saturation point, where further updates have minimal impact, is reached three times faster in the real-time mode (at $1\, 500$ views) compared to the batch mode (at $3\, 500$ views). These findings highlight the efficiency and accelerated learning capabilities of updating embeddings in real-time compared to the batch approach.

{\bf Validation from user engagement metrics}:
To validate the above insights, we go back and look at the business metrics, such as video click rates and successful video play percentage, for both real-time and batched updates and plot them in Fig. \ref{fig:business_metrics_ctr}.
We see that the real-time model has a much better performance in terms of video click rate and successful video play percentage in the lower views bucket (especially at $\le5\,000$ views). This is because it is able to figure out the optimal content embedding much faster than batch $3 \times$ faster saturation cutoff as shown in Fig. \ref{fig:peak_learning_plot_realtime}). This, in turn, results in better content targeting and higher user engagement on the platform.
\vspace{-2mm}

\subsection{Embedding Norm and Popularity Bias}

Since most ranking approaches use the dot product of user and content embeddings, content embeddings with a higher norm would result in a higher ranked order for that content, thereby recommending it more. Keeping this in mind, we investigate how the $\ell_2$-norm is distributed across content of different view buckets once the embedding matures. In general the norm\_ratio at ${\rm view}=x$ can be defined as:

\begin{equation} \label{eq:2}
    {\rm norm\_ratio}_{\text{view}=x} = \frac{\|\text{emb}_{\text{view=} x}\|_2}{\|\text{emb}_{\text{view=} 0}\|_2}
\end{equation}
\vspace{0.5mm}

In Figures \ref{fig:norm_ratio_realtime} and \ref{fig:norm_ratio_batch}, we plot the histogram of the $\ell_2$-norm distance of content embeddings using Eq \ref{eq:2} 
(replacing x with $100\, 000$) for both real-time and batched trained embeddings, respectively. We observe that the $\ell_2$-norm of the embeddings grows drastically with more views for batched updates. This implies that the model in this case ranks content with more views higher, resulting in higher popularity bias, thereby resulting in the serving of content that is generic and not catering to users’ specific interests leading to loss in user engagement. A possible hypothesis is, in batch learning the accumulation of updates over multiple batches can result in larger changes to the embeddings' norms, especially if the model encounters diverse or conflicting patterns in the data. In real-time training, since the updates are applied more frequently and are based on smaller amounts of data, the changes to the embeddings' norms may be less pronounced compared to batch mode learning.

\begin{figure}[!t]
\begin{minipage}[b]{1.0\linewidth}
    \includegraphics[width=8cm, height=5cm]{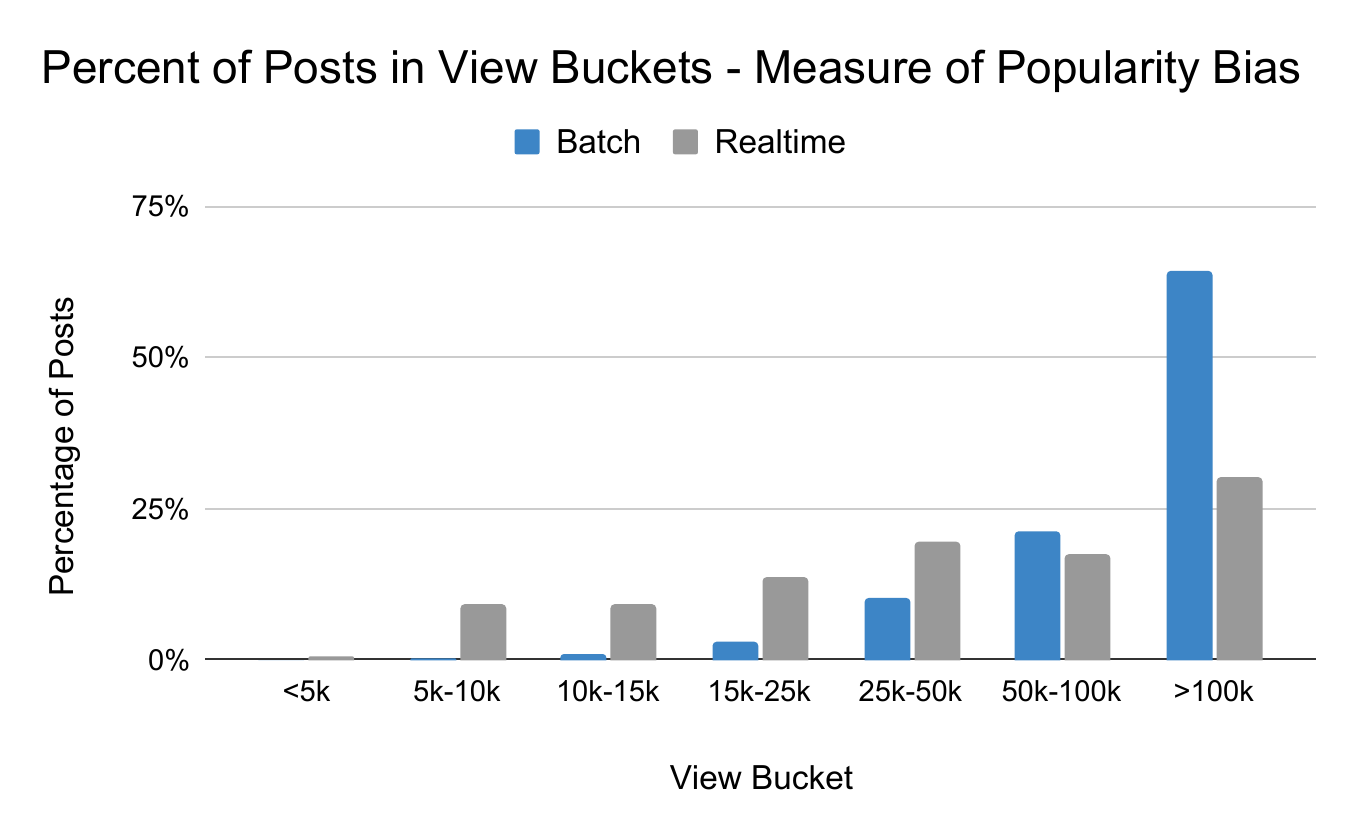}
    \vspace{-4mm}
\end{minipage}
\caption{Prevalence of bias among various view buckets across batch and real-time training methods.}
\label{fig:pop_bias_business}
\end{figure}

{\bf Validation from user engagement metrics}:
To validate the above hypothesis, we plot the business metrics in Fig. \ref{fig:pop_bias_business} that relate to the percentage of views in low (and high) view ranges. We note that the batch model suffers from severe popularity bias with popular posts garnering more than 60\% of the views.

\section{Conclusion \& Future Work}
This paper delves into the intricacies of machine learning production pipelines, explicitly focusing on large-scale recommendation systems. By addressing the complexity and challenges associated with quantitatively monitoring recommendation quality, we have made significant progress in understanding the evolution of embeddings on Sharechat, a social media platform with a substantial user base. Further, we analyzed the business metrics such as successful video views and popularity bias, and show that the proposed metrics for monitoring embeddings highly correlate with the business outcomes. While the analysis is closely linked to the specific algorithm used, the techniques for updating behavioral embeddings remain the same, suggesting generalizability across different recommendation systems and application areas.

For future work, we propose several avenues for more in-depth exploration. One area of interest is determining the minimum number of views necessary to comprehend the complexity levels associated with video or image content on social media. Additionally, we can explore techniques to reduce the frequency of updates based on improvements in maturity to bring down server costs and propose initialization schemes to expedite the maturity process. Finally, the question of whether all user interactions contribute equally to learning is worth looking into, prompting the need to investigate optimal strategies for exposing content to specific user groups to maximize learning outcomes. 

\section{Presenter Bio}
Srijan Saket currently works as a Staff Machine Learning Engineer at \href{https://en.wikipedia.org/wiki/ShareChat}{ShareChat}, developing scalable and cost-effective recommender systems. As a pioneering member of the AI team, he played a pivotal role in establishing the framework for machine learning at ShareChat. Over the course of six years, he worked on transitioning ML projects from research to production, handling tasks such as automated content moderation, creating recommender systems for new categories, and developing scalable, efficient feature pipelines for ranking models. His efforts significantly contributed to the platform's growth, expanding its user community from under 1 million to 200 million+ strong user community. His current research focuses on early stage recommendation, content journey in recommender systems, candidate retrieval for multi-objective ranking, and scalable machine learning systems. Srijan has completed his bachelors and masters from \href{https://www.iitk.ac.in/}{IIT Kanpur} and some of his recent work has been included at top conferences including WWW and RecSys. He delivered the keynote speech (Industrial track) in the recently held \href{http://fire.irsi.res.in/fire/2023/home}{FIRE 2023} conference in Goa, IN. He also holds a granted US patent for human-assisted chatbot conversations. \href{https://www.linkedin.com/in/srijansaket/}{LinkedIn}. \href{https://scholar.google.com/citations?user=hheK6oIAAAAJ&hl=en}{Google Scholar}.

\bibliographystyle{ACM-Reference-Format}
 \balance
\bibliography{sample-base}

\end{document}